\documentclass{andp2012}
\usepackage[english]{babel}

\title{Bound states and Cooper pairs of molecules in 2D optical lattices bilayer}

\author[A. Camacho-Guardian]{A. Camacho-Guardian\inst{1}}
\author[G.A. Domiguez-Castro]{G.A. Dom\'{\i}nguez-Castro\inst{1}}
\author[R. PAredes]{R. Paredes\inst{1}\footnote{Corresponding author\quad E-mail:~\textsf{rosario@fisica.unam.mx}}}
\address[1]{Instituto de F\'{\i}sica, Universidad
Nacional Aut\'onoma de M\'exico, Apartado Postal 20-364, M\'exico D.
F. 01000, Mexico}

\shortauthors{A. Camacho-Guardian et al.}

\begin{abstract}
We investigate the formation of Cooper pairs, bound dimers and the dimer-dimer elastic scattering of ultracold dipolar Fermi molecules confined in a 2D optical lattice bilayer configuration. While the energy and their associated bound states are determined in a variational way, the correlated two-molecule pair is addressed as in the original Cooper formulation. We demonstrate that the 2D lattice confinement favors the formation of zero center mass momentum  bound states. Regarding the Cooper pairs binding energy, this depends on the molecule populations in each layer. Maximum binding energies occur for non-zero (zero) pair momentum when the Fermi system is  polarized (unpolarized). We find an analytic expression for the dimer-dimer effective interaction in the deep BEC regime. The present analysis represents a route for addressing the BCS-BEC crossover superfluidity in dipolar Fermi gases confined in 2D optical lattices within the current experimental panorama.
\end{abstract}
\shortabstract

\begin{document}
\maketitle

\section{Introduction}
\label{intro}
Recent advances in experimental research on cooling and trapping macroscopic samples of ultracold molecules represent the main support for theoretically investigating quantum phases, in particular, those resulting as a function of dimensionality and anisotropy. We can mention density ordered \cite{Parish}, Berezinskii-Kosterlitz-Thouless (BKT) \cite{Bereziinskii,Kosterlitz}, Fulde-Ferrell-Larkin-Ovchinnikov-type (FFLO) \cite{Fulde,Larkin} and high $T_c$ superfluid Fermi phase among others \cite{Perali,Chen}. Of particular interest is the experimental realization of fermonic superfluidity in 2D since it represents the quantum simulator of high$T_c$ superfluidity and superconductivity, phenomena that are perhaps among the most fundamental and long standing condensed matter problems. The many-body phases here referred may arise in molecular ultracold gases as a consequence of microscopic few body interactions, namely, the suppression of bimolecular chemical reactions when polar molecules are confined in quasi-2D configurations \cite{Miranda} and the essential mechanism of two-particle quantum collisions that leads either, to binding scattering of pairs or bound dimeric molecules. Nowadays, one of the major successes in a laboratory of quantum matter is the capability of handling externally, both, the two-body interactions and adjusting/tuning light structures where the molecules can move. In this context, it is worth to mention the current experiments performed with confined ultracold $^{23}$Na$^{40}$K  molecules  in which absolute ground state has been achieved \cite{Park}, as well as the experimental realization of ultracold dipolar magnetic atoms in 2D where the interaction can be varied externally \cite{Frisch}. This system is very versatile in the sense that the interaction between molecules can be varied by changing the dipolar moment in each molecule, both in magnitude and direction, via externally applied electric or magnetic fields \cite{Quemener}. 

In this work we address the formation of Cooper pairs, bound states and dimer-dimer elastic scattering of dipolar Fermi molecules confined in a 2D square crystalline environment. For this purpose we consider molecules with its electric dipole moment aligned perpendicular to a double array of square optical lattices layers (see Fig. \ref{Fig1}). This system can be thought as fermions in two different hyperfine states confined in a 2D square optical lattice with an effective interaction given by a dipolar interaction between molecules. We investigate the formation of both, bound molecular states and Cooper pairs \cite{Cooper}, as a function of dipolar molecule interactions maintaining fixed the ratio of the inter-layer spacing to the lattice constant defining the square lattice. The bound dimeric pairs problem is solved with the standard variational method, while the two-molecule scattering problem is analyzed as in the original Cooper situation, considering the formation of a bound pair resulting from the exclusion principle when a degenerated Fermi gas exists. We also investigate the formation of Cooper pairs for unbalanced Fermi energies of each species, molecules lying in layers $A$ and $B$ respectively. As it is well known such binding pairs are the precursors of the FFLO phases \cite{Fulde,Larkin} . 

Previous studies for atoms with contact pseudo-potential interactions confined in a 3D lattice \cite{Martikainen} have shown qualitative and quantitative differences of the Cooper problem with respect to the homogeneous situation, namely, pairs with nonzero center of mass momentum can be favored if the Fermi energy of each "spin" state of fermions differs \cite{Valiente,Wouters,Valiente1}. Here, besides demonstrating these results for 2D lattices, we determined the dependence of the binding energy on the interaction strength of the long range dipolar interaction between molecules. Regarding the scattering of dimers composed of molecules we find an analytic formula to describe dimer-dimer interactions in the deep BEC regime when bound states are already formed. This result is promising in a double manner, on the one hand, current experiments with ultracold dipolar molecules allow for achieving superfluidity of BCS and bound dimers, and on the other hand, BEC-BCS crossover can be addressed theoretically since the terms defining the many-body Hamiltonian are  already set.  There exist  literature concerning both, the study of dipolar gases at zero and finite temperature in homogeneous environments   \cite{Volosniev,Zinner,Pikovski,Potter}, and analysis taking into account lattice confinements \cite{Vanhala,Camacho}. The present analysis dealing with few-body states of fermions is a key ingredient of the many-body systems dealing with the BEC-BCS crossover.

This work is organized in 6 sections. First in section 2 the model is introduced. Then, section 3 deals with the Cooper problem for both cases, balanced and unbalanced Fermi species populations. In sections 4 and 5 we concentrate in the analysis of bound states and their elastic scattering. Finally in section 6 we present a summary of our investigation.

\section{Model}
\label{section1}
\begin{figure}[htbp]
\begin{center}
\includegraphics[width=2.5in]{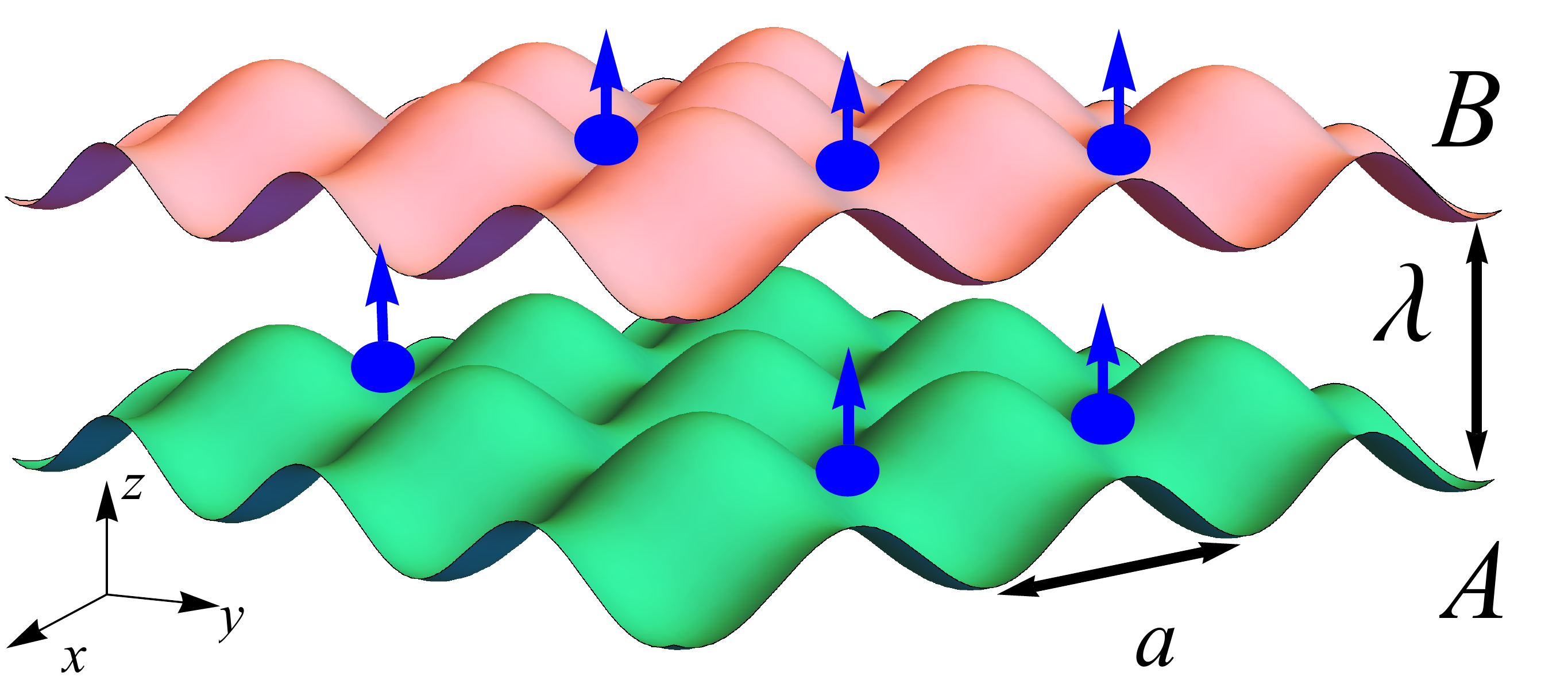} 
\end{center}
\caption{(Color online) Schematic representation of the dipolar Fermi gas. Polar molecules in the up (down) layer can be mapped into the specie labeled with $\uparrow$ ($\downarrow$) when the gas is described in a 2D layer.}
\label{Fig1}
\end{figure}

To address the study depicted above we consider the following model: polar Fermi molecules are placed in a couple of 2D parallel square optical lattices, of lattice constant $a$, with the dipoles aligned perpendicular to the bilayer arrangement (see Fig. \ref{Fig1}). This assumption allows us to neglect the intra-layer interaction and to concentrate in the dominant interlayer interaction\cite{Baranov,Baranov1}. The interlayer interaction is given by
\begin{equation}
V(\vec{x}_A,\vec{x}_B)=d^2\frac{r^2-2\lambda^2}{(r^2+\lambda^2)^{5/2}},
\end{equation}
being $d$  the dipole moment, $\lambda$ the separation among the layers, and $r$ the intra-planar distance $r=|\vec{x}_A-\vec{x}_B|=\sqrt{(x_A-x_B)^2+(y_A-y_B)^2}$, with  $A$, $B$  the labels for Fermi molecules in each layer. Considering this definition, here and hence forth all the distances are considered within a single layer. 
It is convenient to introduce two dimensionless parameters, the scaled separation among layers $\Lambda=\lambda/a$, and the effective interaction strength $\chi = a_{d}/\lambda = m_{eff}d^2/(\hbar^2 \lambda)$, written in terms of the effective mass of an isotropic lattice $m_{eff}=\hbar^2/2ta^2$, being $t$ the hopping strength among nearest neighbors. For current experiments\cite{Park} having $a=532nm$, it is possible to induce dipole moments as large as $0.8D$.  We are considering molecules in its absolute ground state, in those experiments having strong lattice confinement, the recoil energy $E_r= \hbar^2 k^2/2m$  (where $m$ is the molecular mass and $k=\sqrt{k_x^2+k_y^2}$ the wave vector defining the square lattices $V_{latt}(\vec r)= V_0\left(\sin^2(k_x x)+ \sin^2(k_y y)\right)$) is much smaller than the intensity of the optical confinement $V_0$, thus implying that for a separation $\lambda\approx 0.75 a$ one obtains $0.4<\chi<2.0$. All of our calculations were performed taking into account these values for the parameters.

\section{Cooper problem}
\label{section2}

Schr\"odinger equation for the two body wave function $\Phi(\vec{x}_A,\vec{x}_B)$ is,
\begin{equation}
\left(\hat{T}_A+\hat{T}_B+{\hat V}(\vec{x}_A,\vec{x}_B)\right)\Phi(\vec{x}_A,\vec{x}_B)=E\Phi(\vec{x}_A,\vec{x}_B),
\end{equation}
where $\hat T_\alpha$ is an operator that includes the kinetic energy and the optical lattice confinement. Because of the form for the interaction among fermions of types $A$ and $B$, the two-body Schr\"odinger equation can be rewritten in terms of the center of mass and relative intra-planar coordinates $\vec R$ and $\vec r$ respectively. Considering this observation and following the same reasoning as that in the original Cooper formulation, one can propose a two-molecule wave function as the product $\Phi(\vec{x}_A,\vec{x}_B)=e^{i\vec{K}\cdot\vec{R}}u_{\vec{K}}(\vec{R})\psi(\vec{r})$, being $\vec K$ the center of mass wave vector in the reciprocal space belonging to the first Brillouin zone, $u_{\vec K}(\vec R)$ a periodic function having half period of the square lattice and $\psi(\vec r)$ a relative wave function to be determined. In this ansatz we have assumed both, that the lattice is deep enough to consider the lowest band only and that the tight binding approximation holds. It is important to note that because of the distinguishability of the molecules $A$ and $B$ it is not necessary to write the antisymmetric wave function for the pair. It is straightforward to show that the equation that $\psi(\vec{r})$ satisfies is,
\begin{equation}
(\vec{\xi}_{\vec{K}}\cdot\vec{\hat{T}}_D+V(\vec{r}))\psi(\vec{r})=E\psi(\vec{r}).
\label{K-r}
\end{equation}
where $\vec{\xi }_{\vec{K}}=-2t(\cos(K_x a/2),\cos(K_y a/2))$ and $\vec{\hat{I}}\cdot\vec{\hat{T}}_D\psi(\vec{r})=\sum_{i=x,y}\left(\psi(\vec{r}+\vec{\delta}_i)+\psi(\vec{r}-\vec{\delta}_i)\right)$, where $\vec{\delta}_i=a\hat{e}_{i}$, $\vec{\hat{I}}$ is the $2\times 2$ identity matrix and $\hat{e}_{i}$ the unit vector along the $i$ direction.
Expanding the wave function for the relative coordinate $\psi(\vec r)$ in terms of the relative momenta $ \vec q$
\begin{equation}
\psi(\vec{r})=\frac{1}{N_x N_y}\sum_{\vec{q}}\psi(\vec{q})e^{i\vec{q}\cdot\vec{r}},
\end{equation}
and defining 
$$E_{\vec{K},\vec{q}}=-4t\left(\cos(K_xa/2)\cos(q_xa)+\cos(K_ya/2)\cos(q_ya)\right)$$ one ends with Eq. (\ref {K-r}) written in the momentum representation, 
\begin{equation}
(E-E_{\vec{K},\vec{q}})\psi(\vec{q})=\sum_{\vec{q'}}V(\vec{q}-\vec{q'})\psi(\vec{q'}), 
\end{equation}
Here it is important to stress the qualitative difference of the present study with respect to the original homogeneous Cooper 3D situation\cite{Cooper}. Namely, that the dependence of the center of mass and the relative wave vectors can never be decoupled when the pair is scattered in the presence of the lattice. In view of the last equation we arrive to an equation for the energy $E$ that reads,
\begin{equation}
1=\frac{1}{N_x N_y}\sum_{\vec{q'}} \frac{V(\vec{q}-\vec{q'})}{E_{\vec{K},\vec{q}}-E},
\end{equation}
where the prime in the momentum $\vec{q}$ indicates that the sum has to be restricted to the allowed states. These values of $\vec q$ are, as in the Cooper problem, associated to a pair of molecules interacting in the presence of a quiescent Fermi sea. Thus, the amplitudes for states already occupied are forbidden as a consequence of the exclusion principle. This means that in the sum we should include states, vectors with relative momentum $\vec{q}$, that simultaneously satisfy

\begin{equation}
\epsilon_{\frac{\vec{K}}{2}+\vec{q}}>\epsilon_F^A  \>\>\> \text{and} \>\>\> \epsilon_{\frac{\vec{K}}{2}-\vec{q}}>\epsilon_F^B
\end{equation}
In accordance with the above argument we look for solutions $\psi(\vec r)$ having energies $E=\epsilon_F^A+\epsilon_F^B-\Delta$, with $\Delta>0$, that is, with less energy that the sum of Fermi energy of each layer. Investigation of these energies lead us to determine the energy gap, for different values of the parameter that characterizes the interaction, as a function of the center of mass momentum vector $\vec K$ and of the Fermi energy in layers $A$ and $B$. All of our numerical calculations were done considering $N_x=N_y=5 \times 10^2$. Numerical calculations with larger values of sites produce the same results.

\begin{figure}[htbp]
\begin{center}
\includegraphics[width=3.5in]{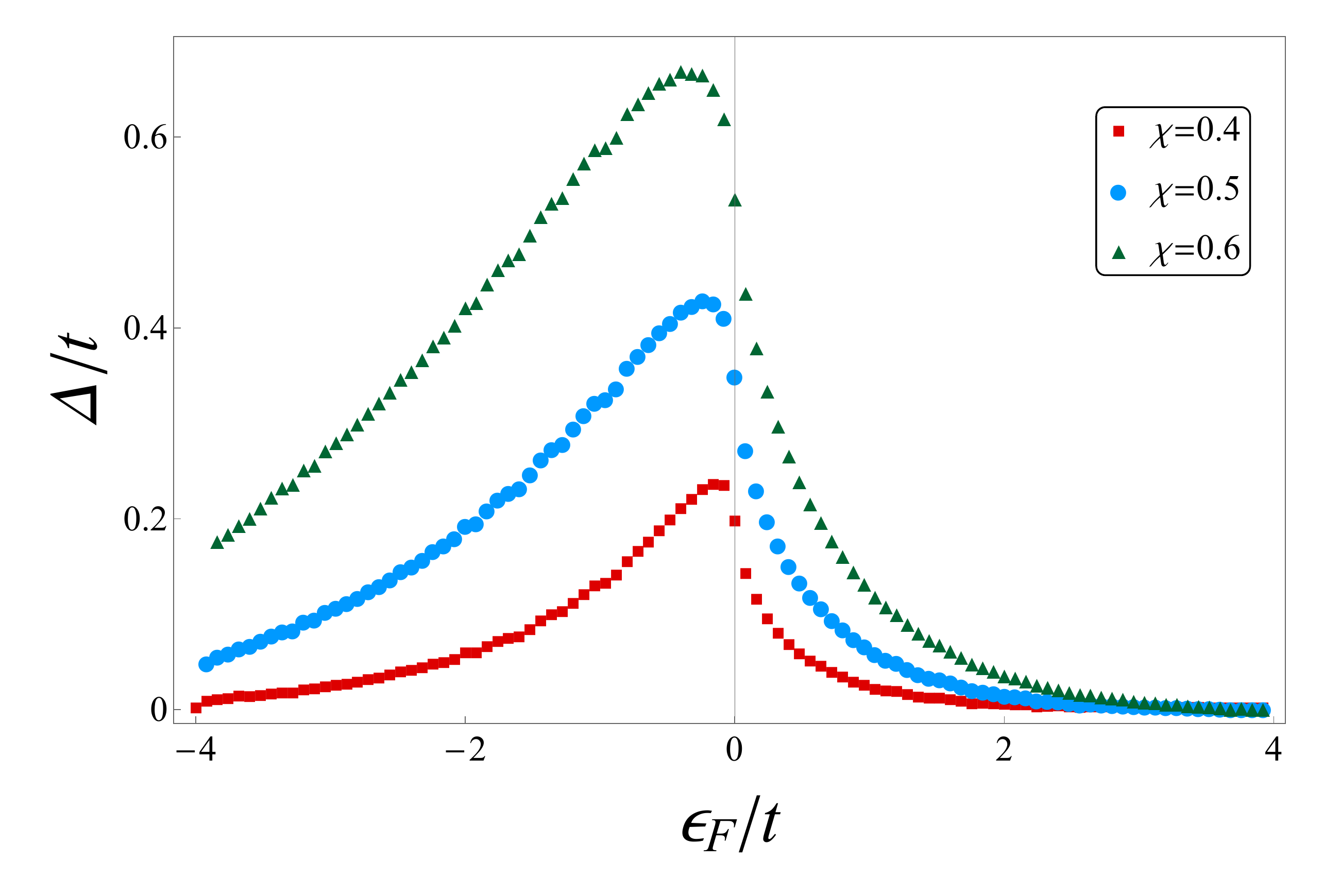} 
\end{center}
\caption{  Energy gap $\Delta$ as a function of the Fermi energy in a symmetric lattice $\epsilon_F^A = \epsilon_F^B$}
\label{gap}
\end{figure}

First, we concentrate in studying the dependence of the energy gap as a function of the Fermi energy for symmetric lattices. Namely $\epsilon_F^A = \epsilon_F^B$. In Fig. \ref{gap} we plot $\Delta$ as a function of the Femi energy, considering several values of $\chi$ for a fixed value of the interlayer separation. For simplicity we assume $K_x=K_y=0$. Squares, circles and triangles, correspond to $\chi=0.4, 0.5$ and 0.6 respectively. As in the study performed for 3D square lattices \cite{Martikainen}, we found that the binding energy increases and reaches a maximum value as a function of the Fermi energy. Then, when Fermi surfaces change from close to open the binding energy shows an exponential decay of the form $\Delta \sim \exp^{-\epsilon_F/\alpha}$, being $\alpha \sim \chi$. We also notice that as the interaction strength increases the binding energy has a maximum value that becomes shifted as the Fermi energy decreases. One can interpret this behavior as a result of the competition among the saturation of the filling factor against the interaction strength, which when increasing together produce an insulating phase.

\begin{figure}[htbp]
\begin{center}
\includegraphics[width=3.5in]{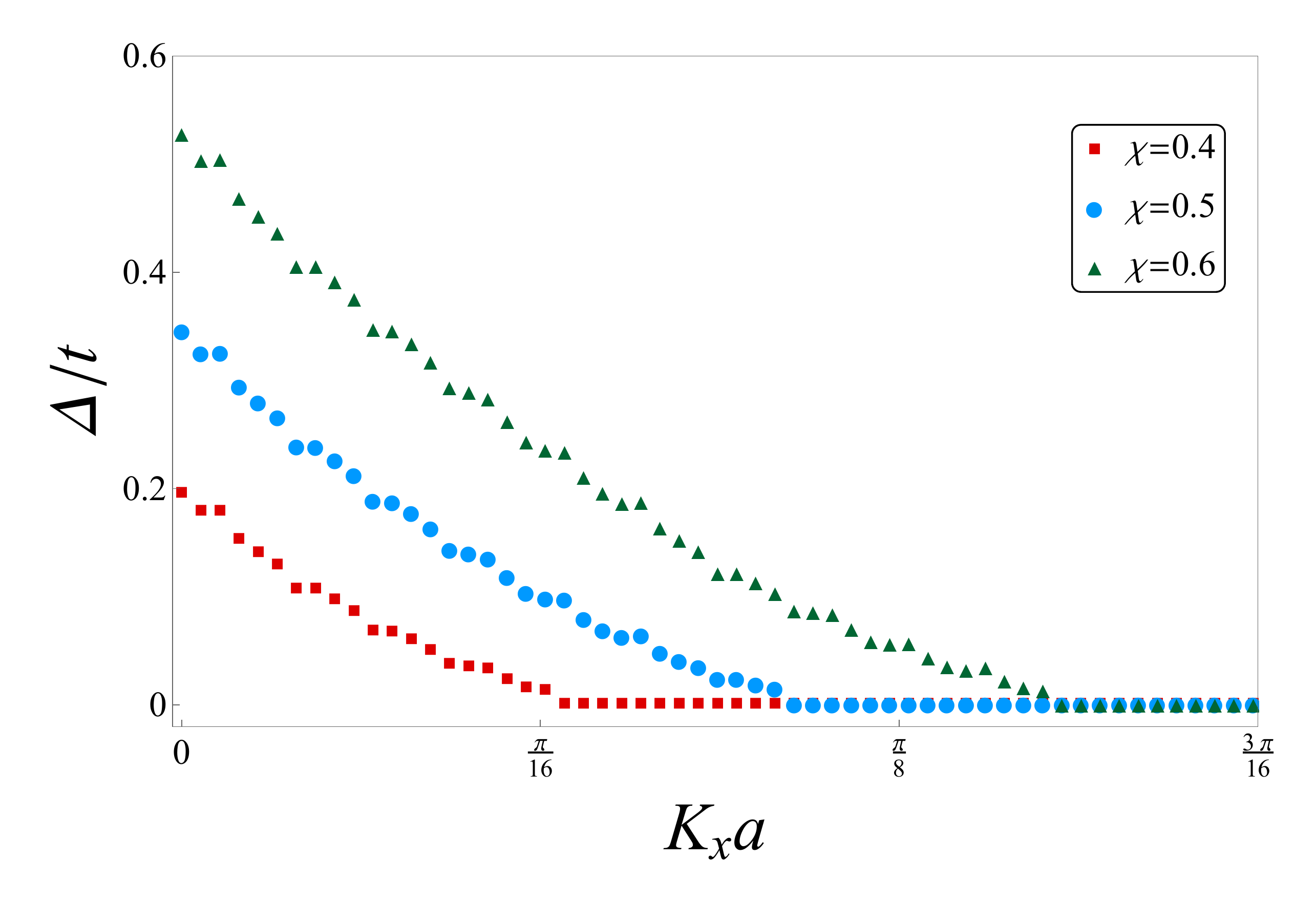} 
\end{center}
\caption{$\Delta$ as a function of $K_x a$ for symmetric Fermi energies $\epsilon_F^A=\epsilon_F^B$. We selected $K_y=0$. From top to bottom $\chi = 0.4, 0.5$ and $0.6$.}
\label{gap_Kx}
\end{figure}

Next, for illustrating the influence of the energy gap on the center of mass momentum $\vec K$, in Fig. \ref{gap_Kx} we plot the dimensionless energy gap $\Delta/t$ as a function of $K_x a$ assuming $K_y=0$ for $\chi$ ranging in the interval $[0.4,0.6]$, considering similar circumstances as those in Fig. \ref{gap} ($\epsilon_F^A = \epsilon_F^B$). We observe that the maximum value of $\Delta$ occurs for center of mass momentum near ${\vec K}=0$. As before, the largest the effective interaction is, the greater the binding energy become. 

\begin{figure}[htbp]
\begin{center}
\includegraphics[width=3.5in]{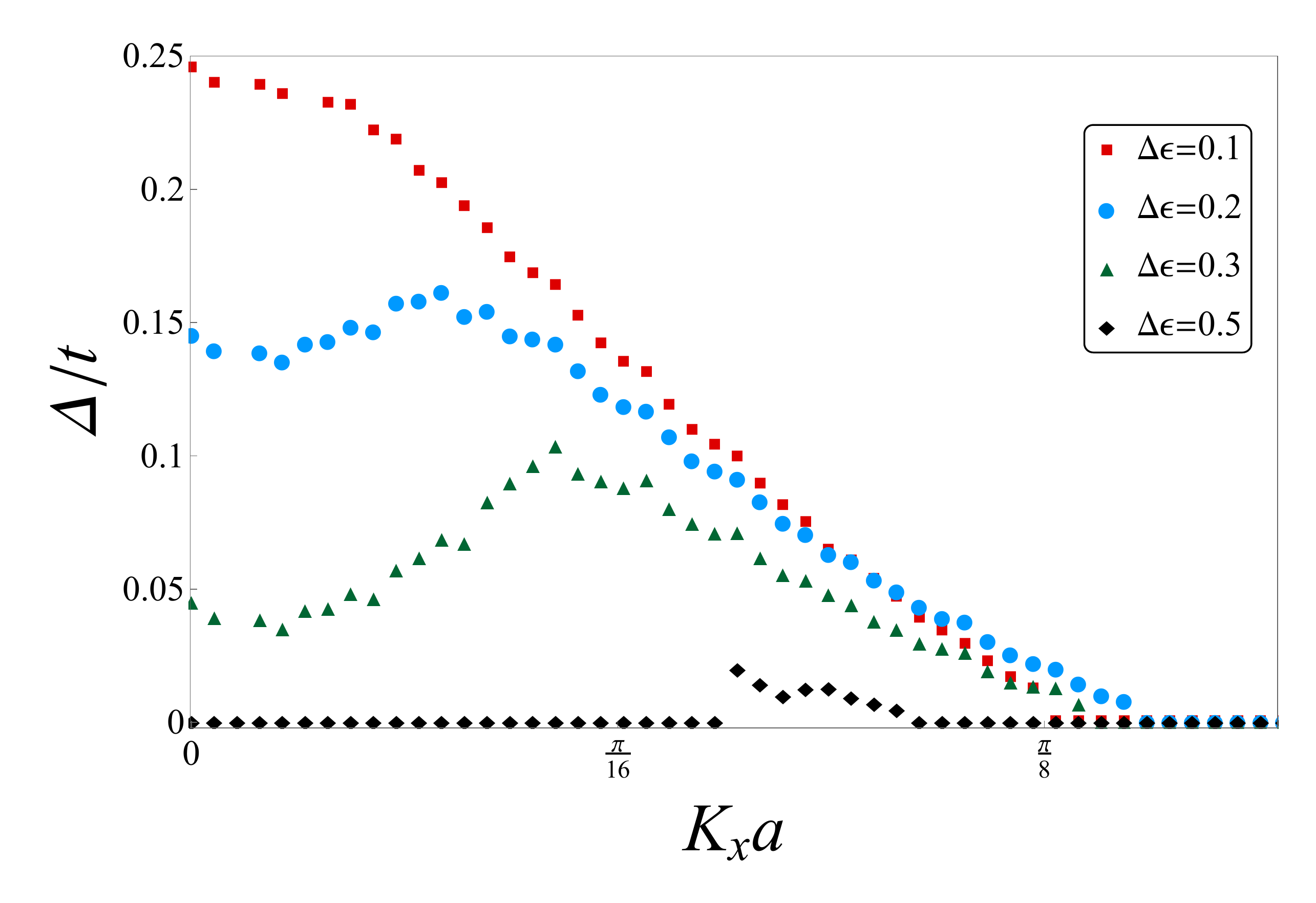} 
\end{center}
\caption{$\Delta$ as a function of $K_x a$ for non-symmetric Fermi energies. We selected $K_y=0$ and $\chi=0.5$. From top to bottom $\Delta \epsilon = 0.1, 0.2, 0.3$ and 0.5.}
\label{gap_Kx-a}
\end{figure}

To complement the study of the Cooper problem we analyze the dependence of the energy gap on the center of mass momentum for non-symmetric values of the Fermi energy in layers $A$ and $B$. To exemplify that formation of pairs also happens in this case, we select $\chi=0.5$ and determine the behavior of $\Delta$ for unbalanced Fermi energies $\Delta  \epsilon = \epsilon_A-\epsilon_B$. In Fig. \ref{gap_Kx-a} we plot several curves of the binding energy associated to unequal Fermi energies $\Delta \epsilon = 0.1, 0.2, 0.3, 0.5$. We find, as in the previous study for two-body contact interactions in 3D \cite{Martikainen}, that in systems with unequal Fermi energies formation of Cooper pairs of nonzero center of mass momentum is favored. We also reach the conclusion that when the difference among the Fermi energies $\epsilon_F^A$ and $\epsilon_F^B$ is too large, the Cooper pair cannot be formed. One can explain this behavior as caused by the imbalance in the filling factors which results in the impossibility of pairing molecules of type $A$ lying on the surface with other molecules having largest Fermi energy. However, by properly adjusting the populations of molecules in each layer, the binding energy can be maximized and therefore FFLO phases can be observed.  

\begin{figure}[h]
\begin{center}$
\begin{array}{cc}
\includegraphics[width=1.4in]{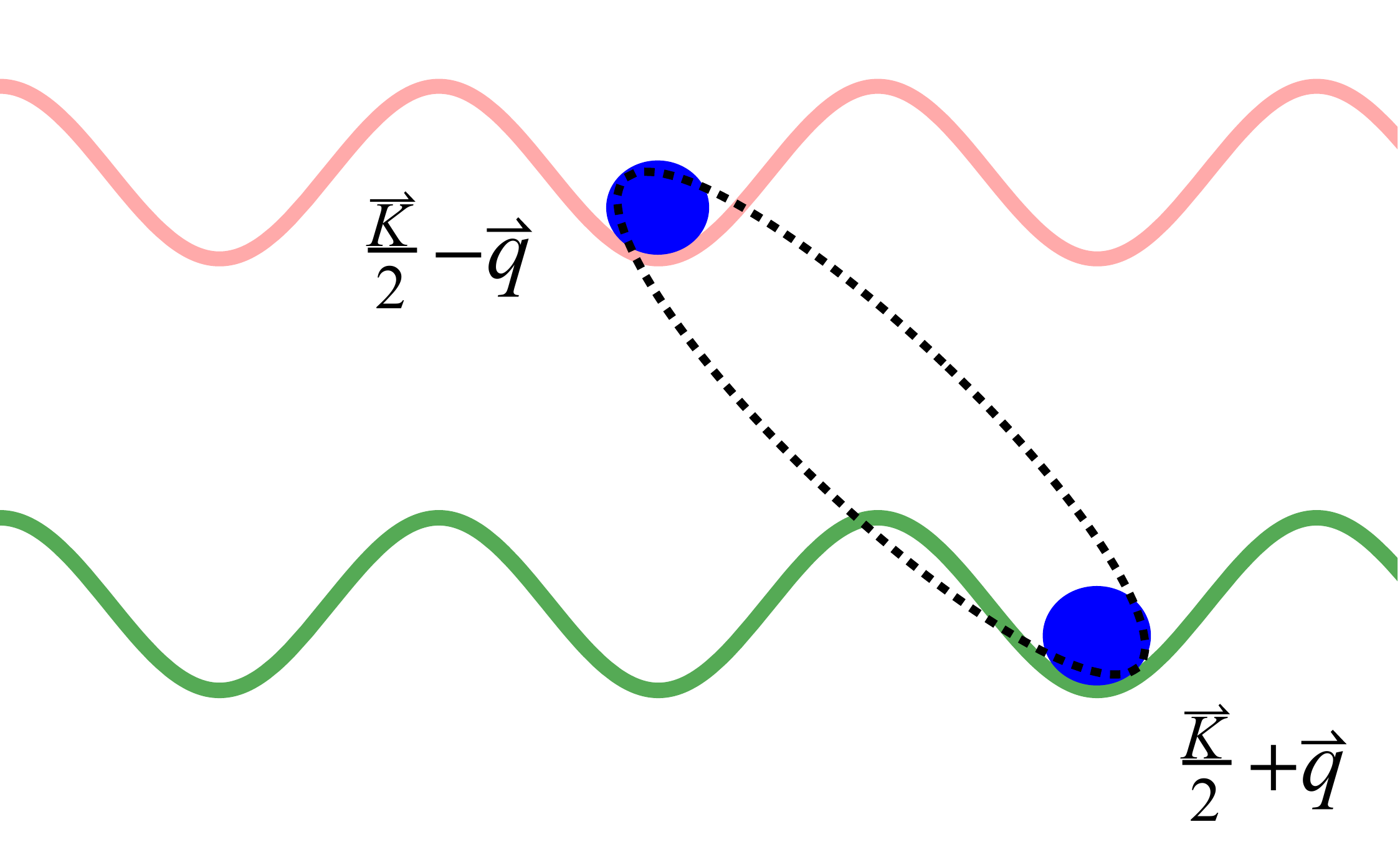} &
 \includegraphics[width=1.4in]{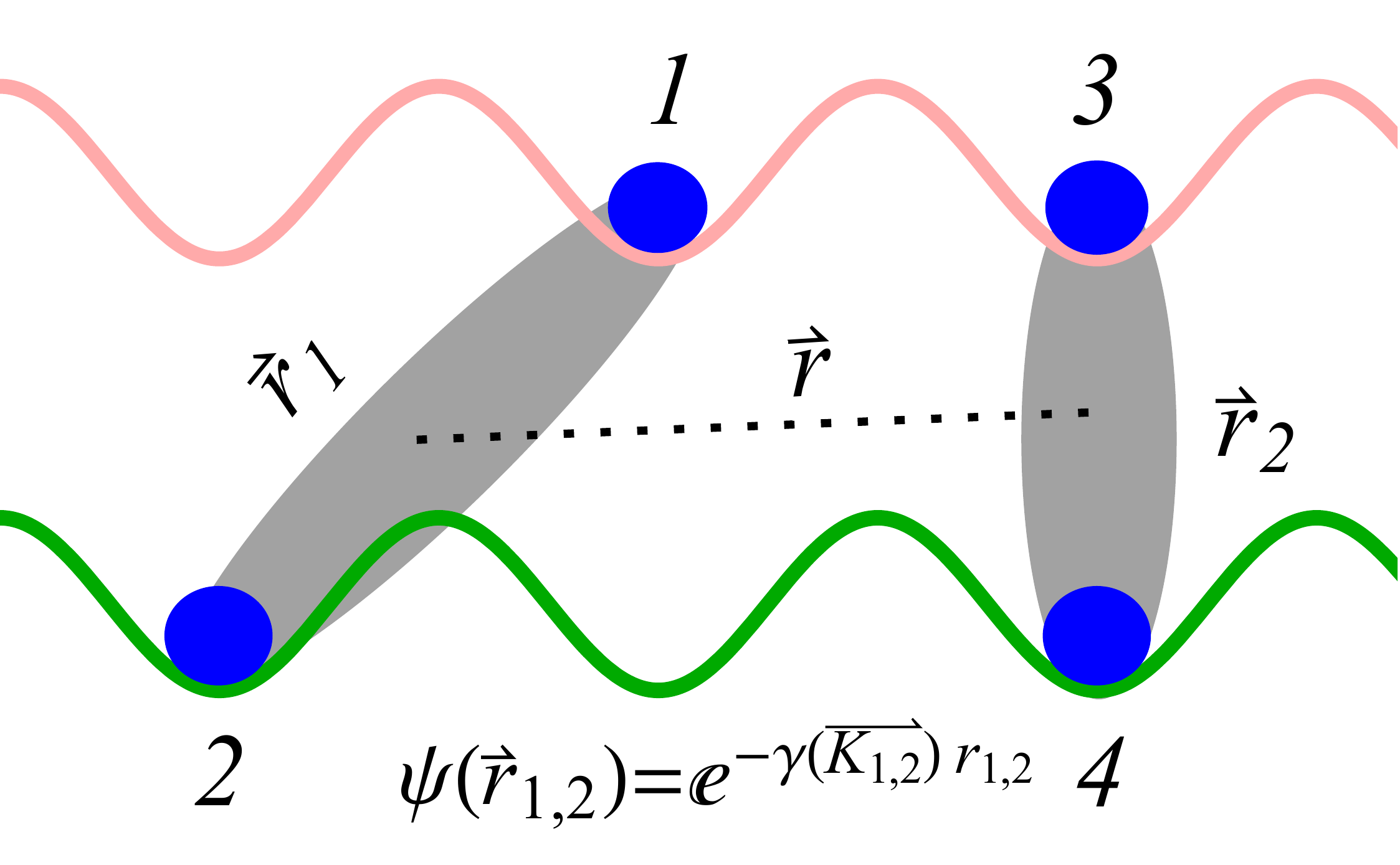}
\end{array}$
\end{center}
\caption{We illustrate in this scheme the possible Cooper pairs (dotted lines) and dimers that can be formed in the bilayer array. On the left Cooper pairs are shown, while on the right shadow ellipses illustrate a couple of bound dimers of molecules.} 
\label{cartoon}
\end{figure}

In Fig. \ref{cartoon} on the left we illustrate the possible pairs that can be formed in our model. We should stress that the for typical experiments with ultracold molecules, confining them in square optical lattices in 2D, this cartoon represent plausible size of pairs. It is interesting to note functional integral calculations of the BCS state at $T=0$ \cite{Camacho} predicts similar values for the binding energy than those here obtained for the two-body problem.

\section{Bound states of molecules}
\begin{figure}[htbp]
\begin{center}
\includegraphics[width=3.5in]{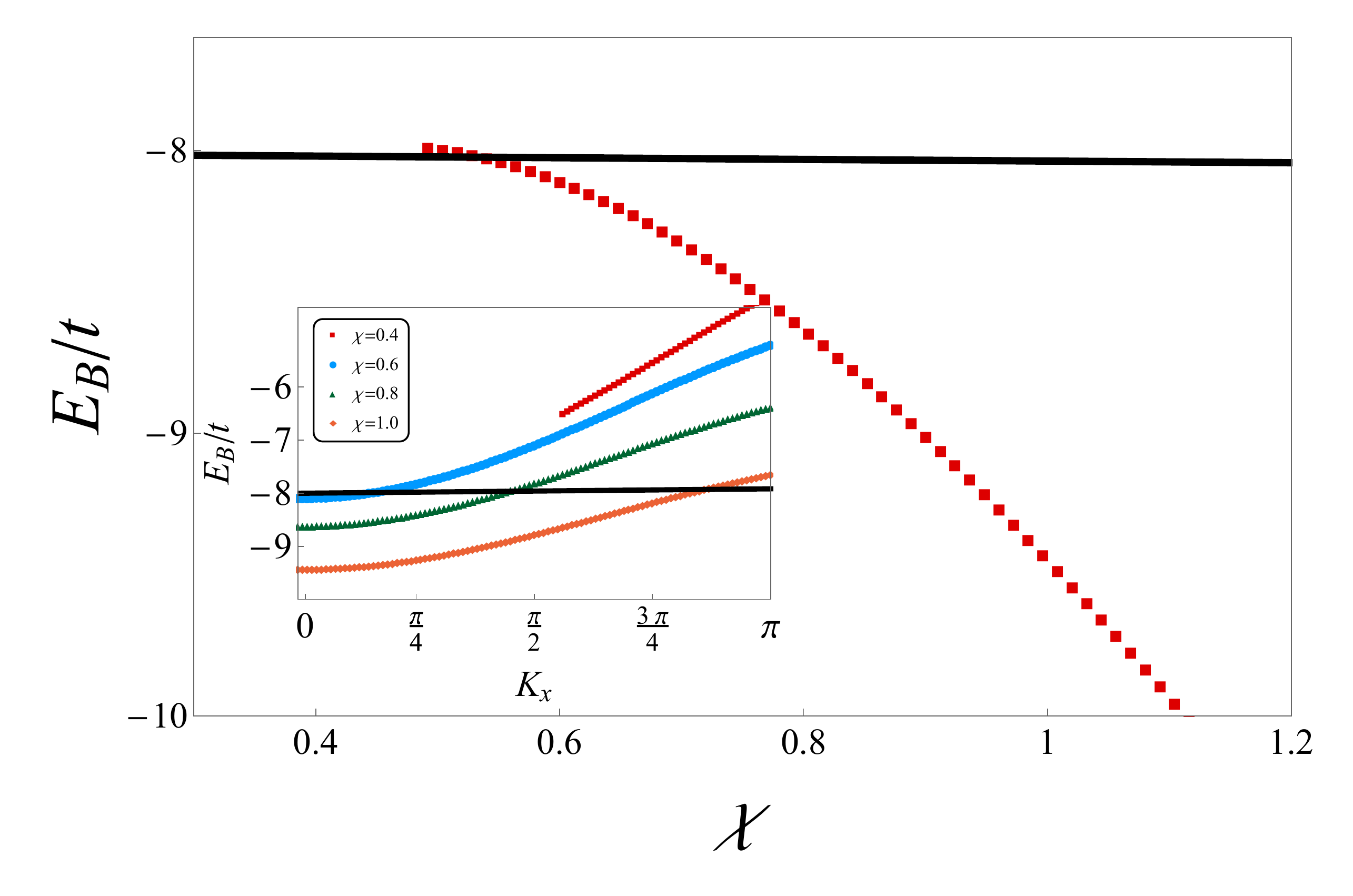} 
\end{center}
\caption{Energy of bound states $E_{B}$ as a function of $\chi$. In the inset we plot curves of $E_B$ as a function of the center of mass reciprocal vector along $x$ direction for $\chi=0.4,0.6,0.8.1.0$.}
\label{EB}
\end{figure}

Now we investigate the existence of true bound states in the square lattice, namely, molecular bound states which are dimers composed of two molecules. To determine the energy of the dimer we use the standard variational method, proposing a two molecule wave function describing the bound pair. We investigate this energy taking into account the dependence on the effective interaction between molecules measured in terms of the parameter $\chi$, for a fixed value of the bilayer array $\lambda = 0.75 a$. The ansantz for the two body problem is
\begin{equation}
\Phi_B(\vec{x}_A,\vec{x}_B)=e^{i\vec{K}\cdot\vec{R}}\psi(\vec{r})
\end{equation}
with $\psi(\vec{r})=Ae^{-\gamma r}$, being $\gamma$ a variational parameter and $A$ a normalization constant. $\vec K$ is the reciprocal center of mass wave vector in the first Brillouin zone. We numerically solve the equation $H \Phi (\vec x_A, \vec x_B)= E_B \Phi (\vec x_A, \vec x_B)$. As in the case of Copper problem our calculations were done considering $N_x=N_y= 250$. In Fig. \ref{EB} we plot the energy as a function of $\chi$. We notice that solutions start to appear from a particular value of $\chi$, that is, for smaller values of $\chi \approx 0.49$ the molecules form a scattered pair. The inset of this figure includes several curves of $E_B$ as a function $K_x$ for $\chi =0.4, 0.6,0.8,1.0$. As can be appreciated from this figure, the trial wave function that we choose does not allow the formation of bound states for arbitrary values of $\chi$. Because of the symmetry of the square lattice, analogous results for $E_B$ as a function of $K_y$ are obtained. As Fig. \ref{EB} demonstrate, formation of bound states is favored for the center of mass $\vec K =0$ as in the unpolarized Fermi system.  

\section{Scattering of bound molecules}
\begin{figure}[htbp]
\begin{center}
\includegraphics[width=3.2in]{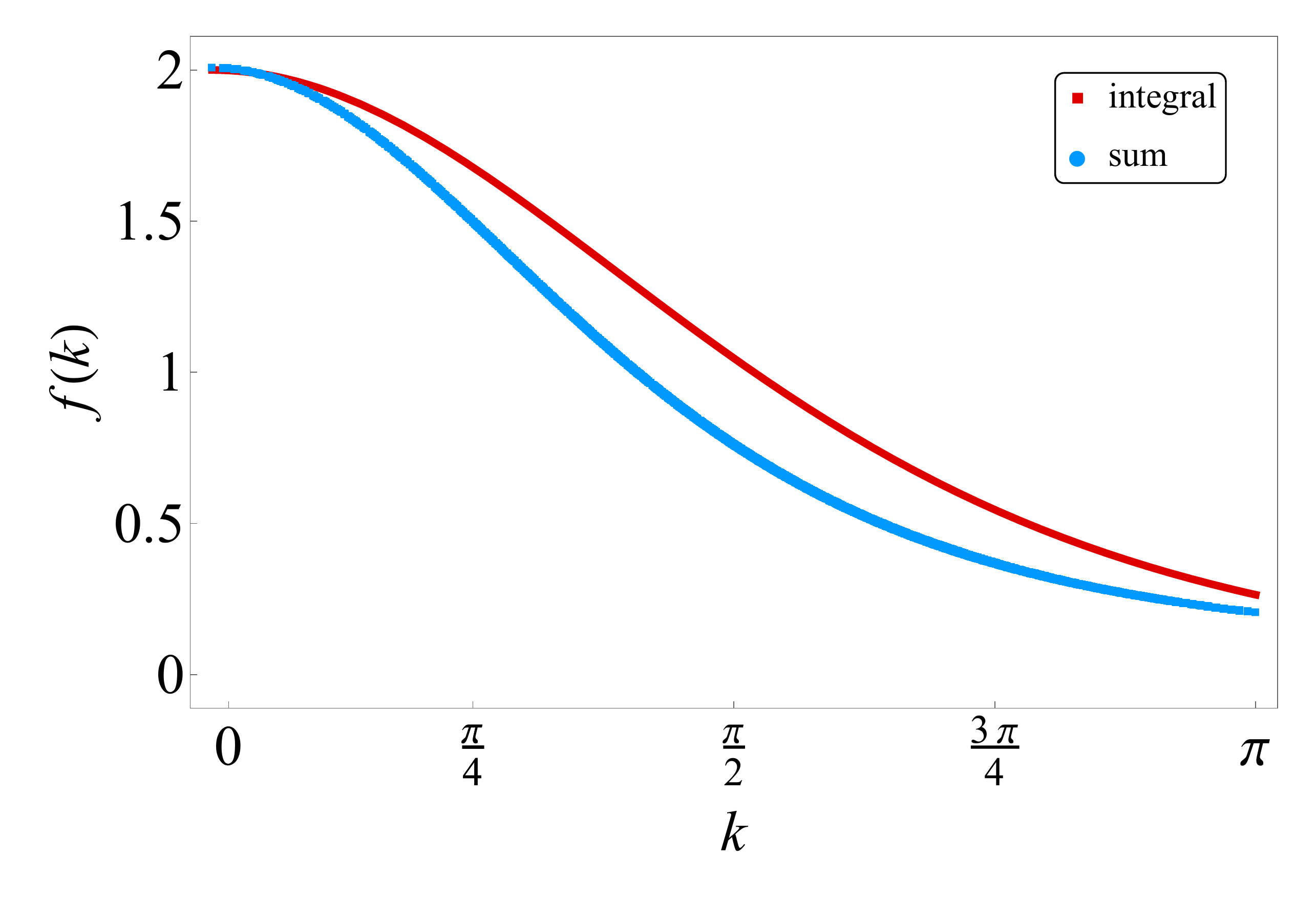} 
\end{center}
\caption{We illustrate the ratio among of the effective interaction potential $V_{eff}(\vec{q}-\vec{q}')$ in Eq. (\ref{Veff}) to the dipolar interaction potential between molecules $V_{dip}(\vec{q}-\vec{q}')$.}
\label{Interaction-Ratio}
\end{figure}

To complete the analysis of the problem for the bound states we now investigate the elastic scattering of interacting pairs, that is, the scattering among dimers composed of two molecules. In Fig.\ref{cartoon} we show schematically two bound pairs of molecules lying in the bilayer array. The molecules composing those bound pairs are labeled with numbers 1 and 2, and 3 and 4 respectively. We assume that these dimers form a couple of rigid dumbbells with a bound energy $E_B$ and a wave function  $\phi_B(\vec{x}_A,\vec{x}_B)$. The dumbbells do not have vibrational nor rotational degrees of freedom. Thus, although molecules interact each other, bound pairs tunnel across the dumbbell jointly. The four-body Hamiltonian of such a system is,

\begin{equation}
H= \sum _{s=1} ^4\hat{T}_s  + {\hat V}(\vec{x}_1,\vec{x}_2)+ {\hat V}(\vec{x}_3,\vec{x}_4)+{\hat V}(\vec{x}_1,\vec{x}_4)+{\hat V}(\vec{x}_3,\vec{x}_2),
\end{equation}   
where the operator ${\hat T}_i$ contains, as before, the kinetic energy and the lattice confinement potential. We look for solutions $\hat H \phi(\vec x_1, \vec x_2, \vec x_3, \vec x_4)= E \phi(\vec x_1, \vec x_2, \vec x_3, \vec x_4)$. For this purpose we first define the center of mass and relative coordinates in terms of $\vec x_i$, $i=1,.. , 4$, as follows, 

\begin{equation}
\vec r_1= \vec x_1- \vec x_2, \>\> \vec r_2 = \vec x_3 - \vec x_4,\>\> \vec r= \vec R_1 - \vec R_2, \>\> \vec R = \frac{\vec R_1 +\vec R_2}{2}
\end{equation}
being $ \vec R_1= (\vec x_1+\vec x_2)/2$ and $ \vec R_2= (\vec x_3+\vec x_4)/2$. The wave function $\phi$ is symmetric with respect to the permutation of the dimers and antisymmetric with respect to permutations of molecules within the same lattice. We proceed in the same way as in the Cooper problem expressing the Schr\"odinger equation in terms of this coordinates, considering tunneling of nearest neighbor only. The ideal term in this case becomes: 

\begin{eqnarray}
&&\sum_{s=1}^{4} \hat T_s \phi \left (\vec r_1, \vec r_2, \vec r, \vec R \right)=\\
& =& -J  \sum _{i=x,y} \phi \left (\vec r_1, \vec r_2, \vec r \pm \vec \delta_i, \vec R \pm \frac{ \vec \delta_i}{2} \right) \nonumber + \phi \left(\vec r_1 \vec r_2, \vec r \mp \delta_i, \vec R \pm \frac{\delta_i}{2} \right) \nonumber \\
 \nonumber\end{eqnarray}
where $J$ is the tunneling coupling constant of dimerized molecules, $J=\frac{t}{2}$. Since the interaction between molecules is independent of $\vec R$, a solution of the form $\phi \left (\vec r_1, \vec r_2, \vec r, \vec R \right)= \Phi(\vec r_1, \vec r_2, \vec r) U_{\vec K} (\vec R) e^{i \vec K \cdot \vec R}$ can be proposed, being $\vec K$ the reciprocal wave vector of the center of mass of the four-body system, and $U_{\vec K} (\vec R)$ a periodic function with half-period of the square lattice. After we substitute this anzats into the four-body Schr\"odinger equation, multiply by $U_{\vec K}^* (\vec R)$, and sum over the vectors $\vec R$ we obtain, 

\begin{eqnarray}
&&E \Phi(\vec r_1, \vec r_2, \vec r) = \\ \nonumber
&&= -2J  \sum _{i=x,y} \cos\left( \frac{K_i a}{2} \right)  \left( \Phi(\vec r_1 , \vec r_2, \vec r \pm \vec \delta_i)+ \Phi(\vec r_1 , \vec r_2, \vec r \mp \vec \delta_i) \right)\\ \nonumber
&+&\left(V(r_1)+V(r_2) \right) \Phi(\vec r_1 , \vec r_2, \vec r)  \\ \nonumber
&+& \left (V \left (\left |\vec{r}+\frac{\vec{r}_1+\vec{r}_2}{2}\right | \right)+V \left (\left |\vec{r}-\frac{\vec{r}_1+\vec{r}_2}{2}\right | \right)\right) \Phi(\vec r_1 , \vec r_2, \vec r).
\label{cuatro}
\end{eqnarray}
In the deep BEC regime, namely when the wave function describing each bound pair is sharply localized, one can write a decoupled solution of the form $ \Phi(\vec r_1 , \vec r_2, \vec r)\approx \psi(r_1)\psi(r_2)\varphi(\vec{r})$, where $ \psi(r_1)$ and $\psi(r_2)$ are the variational wave functions describing the bound pairs determined above. In addition to these substitution, one can write $\varphi(\vec{r})=\sum_{\vec q} \varphi_{\vec{q}} e^{i \vec q \cdot \vec r}$. Then, after summing over the $\vec{r}_1$ and $\vec{r}_2$ coordinates, we obtain the equation for $\varphi_\vec{q}$, 
\begin{eqnarray}\nonumber
E\varphi_{\vec{q}}&=&2\left[\xi_B-4J\sum_{i=x,y}\cos\left(\frac{K_i a}{2}\right)\cos(q_i a) \right]\varphi_{\vec{q}}\\
&&+\sum_{\vec{q}'}\varphi_{\vec{q}'}V_{eff} (\vec q -\vec q')
\end{eqnarray}
where $\xi_B= E_B+8t$ and the effective interaction potential is given by,
\begin{equation}
V_{eff}(\vec{q}-\vec{q}')=\sum_{\vec{r}_1,\vec{r}_2}V(\vec{q}-\vec{q'})\cos\left((\vec{q}-\vec{q}')\cdot\frac{\vec{r}_1+\vec{r}_2}{2}\right) \psi^2(r_1)\psi^2(r_2)
\label{Veff}
\end{equation}
This effective potential comes from the interaction between molecules belonging to different dimers. When the sums over $\vec r_1$ and $\vec r_2$ is replaced by an integral $V_{eff}(\vec{q}-\vec{q}')=2\frac{V(\vec{q}-\vec{q'})}{\left( \left |\frac{\vec{q}-\vec{q'}}{4\gamma(\chi)}\right |^2+1\right)^3}$. Thus, from this expression for the effective interaction between dimers one arrives to the conclusion that the dimer-dimer interaction is screened by a factor of $f(|\vec{q}-\vec{q}'|)=f(k)$ where $f(k)=\frac{2}{\left( \frac{k^2}{4\gamma(\chi)}+1\right)^3}$. In Fig. \ref{Interaction-Ratio} we plot $f(k)$ as a function of $k$, blue and red curves correspond to discrete sums over $\vec r_1$ and $\vec r_2$ and the analytic expression when the sums are replaced by integrals respectively. This behavior is reminiscent of the well known result for contact interactions in 3D homogeneous space \cite{Salomon} where the effective interaction between dimers formed by atoms changes from $2a$ to $0.6a$, being $a$ the $s$-wave scattering length.  

Our analysis for the dimer-dimer elastic scattering leads us to obtain an analytic expression for the effective potential among dimers in the deep BEC regime. Thus, with this information one can write the many-body Hamiltonian describing dimers lying on 2D square lattice arrays, namely,
\begin{equation}
\hat{H}_{d}=\sum_{\vec{k}}(\epsilon_{d}(\vec{k})-\mu)c^\dagger_{\vec{k}}c_{\vec{k}}+\sum_{\vec{k},\vec{k'},\vec{q}}V_{eff}(\vec{k}-\vec{k'})c^\dagger_{\vec{k}}c^\dagger_{\vec{q}-\vec{k}}c_{\vec{q}-\vec{k}'}c_{\vec{k}'}
\end{equation}
where  $\epsilon_d(\vec k)=\xi_B-2J\sum_{i=x,y}\cos(k_i a)$, and the subscript $d$ make explicit the fact that single components are the Bose dimers formed by pairs of molecules \cite{Trefzger,Goral}.

\section{Final remarks}
\label{section3}
We have solved the problems of Cooper pairs, molecular bound states and scattering of bound dimers, of dipolar Fermi molecules confined in a double array of square optical latices in 2D. We determined the influence of both, the finite range interaction potential and the lattice confinement on the optimal binding energy of the Cooper pairs, the bound states or dimers composed of two molecules and the dimer-dimer elastic scattering. Numerical calculations for square lattices of size $N_x=N_y=5 \times 10^2$ allowed us to demonstrate that zero center of mass momentum bound pairs are favored, while being this picture modified in the case of Cooper pairs. Polarized and unpolarized populations in each layer produce maximum binding energies for non-zero and zero pair momentum center of mass pair momentum respectively. We also found that the binding energy of Cooper pairs is maximized as the inter-molecule interaction strength increases. In addition to the two-molecule problem we studied the elastic scattering between dimers formed by two molecules and found a closed formula for the effective dime-dimer interactions in the deep BEC regime were bound pairs are already formed.  

Given the correspondence among two-component Fermi systems confined in 2D crystalline environments and the model here proposed, the present study is promising for achieving superfluidity in 2D lattices within the current experimental research. Even more, considering the possibility of externally modifying the electric field or magnetic field \cite{Park}, to controlling two-molecule interactions, our model provides the specific form that the dimer-dimer interaction must have in the deep BEC to address the entire crossover from BCS to BEC regimes from a theoretical perspective.

{\bf Acknowledgments}

\noindent
This work was partially funded by grant IN107014 DGAPA (UNAM). A.C.G. acknowledges scholarship from CONACYT.




\begin{thebibliography}{0}
  
  \bibitem{Parish}%
   \textsc{M. M. Parish} and \textsc{F.M. Marchetti}, \jr{Phys. Rev. Lett.}  \textbf{108}, 145304 (2012).
  
  \bibitem{Bereziinskii}%
  \textsc{ V. Berezinskii}, 
  \jr{Sov. Phys. JETP}  \textbf{32}, 493 (1971), 
  
  \bibitem{Kosterlitz}
  \textsc{ J. Kosterlitz} and 
  \textsc{D. Thouless}
   \jr{J. Phys. C}  \textbf{6},12 (1973).
  
 \bibitem{Fulde}%
  \textsc{P. Fulde} and 
  \textsc{R. A. Ferrell}, \jr{Phys. Rev.}  \textbf{ 135}, A550 (1964),  
  
  \bibitem{Larkin}
 \textsc{ A. I. Larkin} and 
 \textsc{Y. N. Ovchinnikov}
  \jr{Zh. Eksp. Teor. Fiz.}  \textbf{47}, 1136 (1964).  
 
 \bibitem{Perali}%
\textsc{A. Perali}, \textsc{D. Nelson} and \textsc{A.R. Hamilton}, \jr{Phys. Rev. Lett.}  \textbf{110}, 146803 (2013).
  
      \bibitem{Chen}
    \textsc{Q. Chen},
    \textsc{J. Stajic},
    \textsc{S. Tan} and
    \textsc{K. Levin}
    \jr{Phys. Rep} \textbf{412}, 1 (2005).
  
\bibitem{Miranda} %
 \textsc{M. H. G. de Miranda}, et al.
  \jr{Nat. Phys.}  \textbf{7}, 502 (2011).
 
 \bibitem{Park} 
 \textsc{J. W. Park}, 
 \textsc{S.A. Will} and 
 \textsc{M.W. Zwierlein}
  \jr{Phys. Rev. Lett.}  \textbf{ 114}, 205302 (2015).
  
 \bibitem{Frisch}
\textsc{A. Frisch},
\textsc{M. Mark},
\textsc{K. Aikawa},
\textsc{S. Baier},
\textsc{R. Grimm},
\textsc{A. Petrov},
\textsc{S. Kotochigova},
\textsc{G. Qu\'emener},
\textsc{M. Lepers},
\textsc{O. Dulieu} and
\textsc{F. Ferlaino}
\jr{Phys. Rev. Lett. } \textbf{115}, 203201 (2015).

\bibitem{Quemener}
\textsc{G. Qu\'em\'ener} and
 \textsc{P. Julienne }
 \jr{ Chem. Rev. }\textbf{112} 4949 (2012).

  \bibitem{Cooper}%
  \textsc{L. N. Cooper}, 
  \jr{Phys. Rev.} \textbf{104}, 1189 (1956).
  
  
  \bibitem{Martikainen}%
  \textsc{J. -P. Martikainen}
  \jr{Phys. Rev. A}   \textbf{78}, 035602 (2008).
  
   \bibitem{Valiente}%
  \textsc{M. Valiente} and
  \textsc{D. Petrosyan}, 
   \jr{J. Phys. B.}   \textbf{41}, 161002 (2008).
   

  
  \bibitem{Wouters}%
\textsc{M. Wouters} and
  \textsc{G. Orso}
  \jr{Phys. Rev. A}   \textbf{73}, 012707 (2006).

\bibitem{Valiente1}
\textsc{M. Valiente}
\jr{ Phys. Rev. A} \textbf{81} 042102 (2010).

  \bibitem{Volosniev}
\textsc{A. G. Volosniev},
\textsc{N. T. Zinner},
\textsc{D. V. Fedorov},
\textsc{A. S. Jensen} and
\textsc{B. Wunsch}
\jr{J. Phys. B} \textbf{44} 125301 (2011).

  \bibitem{Zinner}
  \textsc{N. T. Zinner},
  \textsc{J. R. Armstrong},
  \textsc{A. G. Volosniev},
  \textsc{D. V. Fedorov} and
  \textsc{A. S. Jensen}
  \jr{Few Body Syst} \textbf{53} 369 (2012).


    \bibitem{Pikovski} 
    \textsc{A. Pikovski},
    \textsc{M. Klawunn},
    \textsc{G. V. Shlyapnikov} and
    \textsc{L. Santos}
    \jr{Phys. Rev. Lett.} \textbf{105}, 215302  (2010).
    
    \bibitem{Potter}
    \textsc{A. C. Potter},
    \textsc{E. Berg},
    \textsc{D. -W. Wang},
    \textsc{B. I. Halperin} and
    \textsc{E. Demler}
    \jr{Phys. Rev. A} \textbf{105}, 220406 (2010).
    
    \bibitem{Vanhala}
    \textsc{T. Vanhala},
    \textsc{J. Baarsma},
    \textsc{M. Heikkinen},
    \textsc{M. Troyer},
    \textsc{A.  Harju} and
    \textsc{P. T\"orm\"a}
    \jr{Phys. Rev. B} \textbf{91}, 144510 (2015).
    
    \bibitem{Camacho}
    \textsc{A. Camacho-Guardian} and
    \textsc{R. Paredes}
    \jr{Arxiv preprint cond-mat} 1505.03811.
    \url{http://arxiv.org/abs/1505.03811} (2015).

\bibitem{Baranov}
\textsc{M. A. Baranov}
\jr{Phys. Rep} \textbf{464} (2008).

\bibitem{Baranov1}
\textsc{M. A. Baranov},
\textsc{M. Dalmonte},
\textsc{G. Pupillo} and
\textsc{P. Zoller}
\jr{Chem. Rev} \textbf{12} 5012 (2012).
  
      \bibitem{Salomon}
    \textsc{D. S. Petrov},
     \textsc{C. Salomon} and
     \textsc{ G. V. Shlyapnikov}
     \jr{Phys. Rev. Lett.} \textbf{93}, 0900404, (2004).

  
  \bibitem{Trefzger}
  \textsc{C. Trefzeger},
   \textsc{C. Menotti} ,
    \textsc{B. Capogrosso-Sansone} and
     \textsc{M. Lewenstein}
    \jr{J. Phys. B} \textbf{88}, 193001 (2011).
    

     
     \bibitem{Goral}
      \textsc{K. G\'oral}, 
       \textsc{L. Santos} and
        \textsc{M. Lewenstein}
      \jr{Phys. Rev. Lett} \textbf{88}, 170406 (2002).
      


    

\end{thebibliography}

\end{document}